\documentstyle[epsfig,preprint,aps]{revtex}
\tightenlines
\begin{document}
\title{Superheavy Nuclei in a Chiral Hadronic Model}
\author{ Ch. Beckmann, P. Papazoglou, D. Zschiesche\\ 
S. Schramm, H. St\"ocker, W. Greiner\\ {\it Institut f\"ur 
Theoretische Physik\\
Johann Wolfgang Goethe Universit\"at, D-60054 Frankfurt
am Main}\\
}
\maketitle
\begin{abstract}
Superheavy nuclei are investigated in a nonlinear chiral SU(3)-model.
The proton number Z=120 and neutron numbers of N=172, 184 and 198
are predicted to be magic.
The charge distributions and $\alpha$-decay chains hint towards a hollow
stucture. 
\end{abstract}
\begin{figure}[h]
\vspace*{-0.3cm}
\epsfig{file=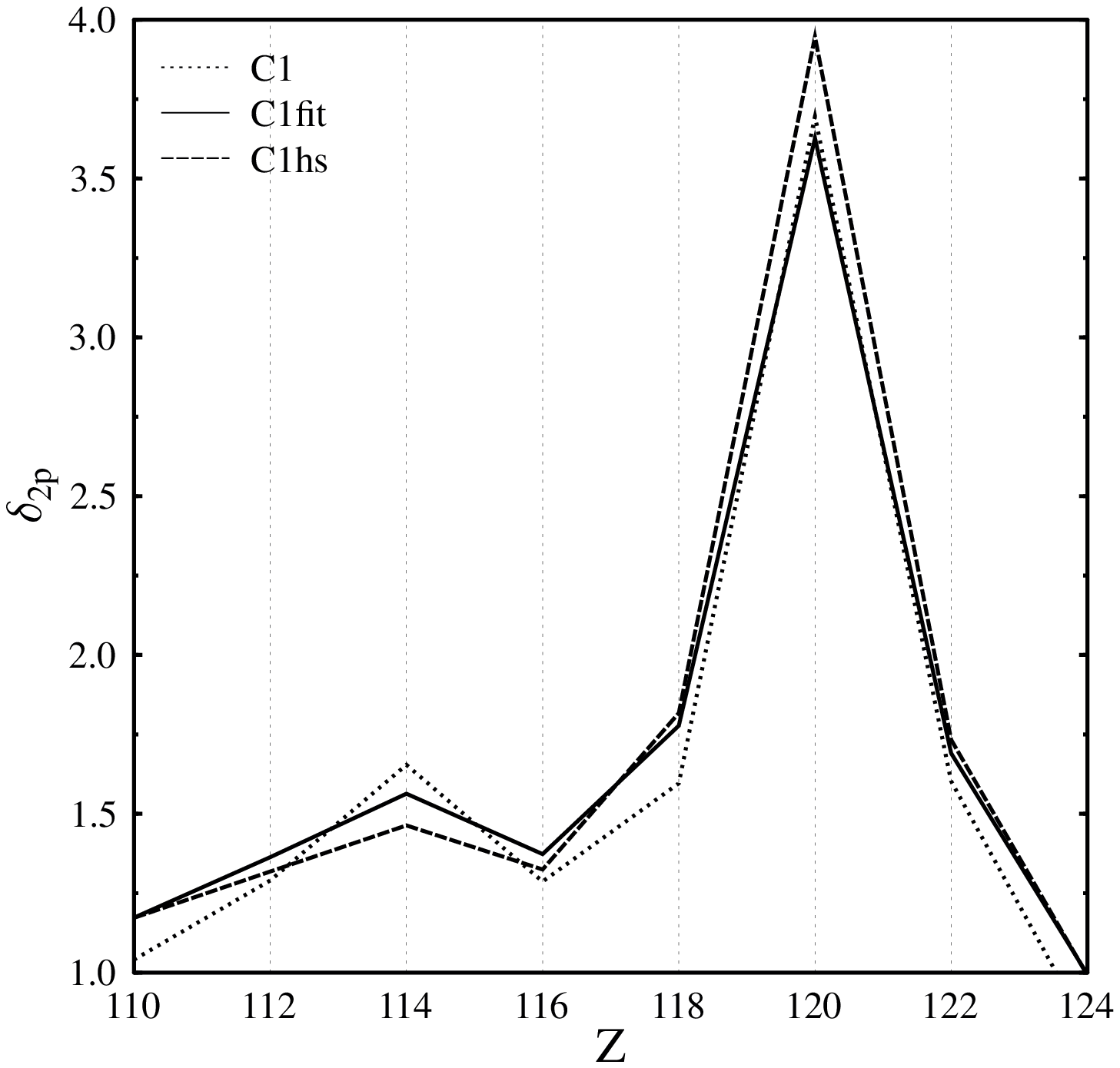, height=10cm ,width=16cm}
\caption{\label{gap}Two-Proton Gap}
\end{figure}
\begin{figure}[h]
\vspace*{-0.3cm} 
\epsfig{file=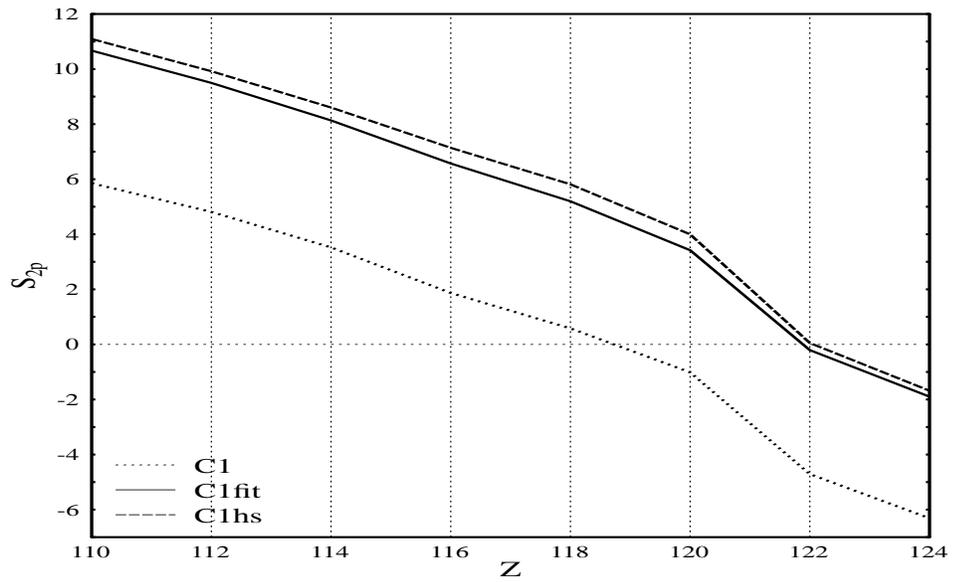, height=10cm ,width=16cm}
\caption{\label{sep}Two Proton Separation Energy $S_{\mathrm 2p}$
for N=172}
\end{figure}
\begin{figure}[h]
\vspace*{-0.3cm} 
\epsfig{file=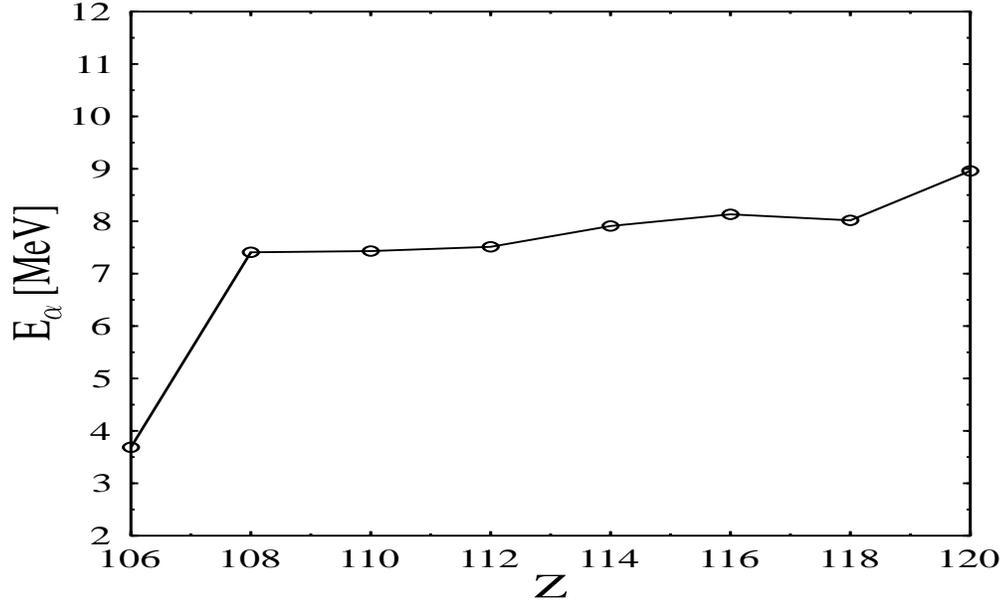, height=10cm ,width=16cm}
\caption{\label{alpha}Predicted Alpha Decay Energy for
Superheavy Nuclei as a Function of Charge}
\end{figure}
\begin{figure}[h]
\vspace*{-0.3cm}
\epsfig{file=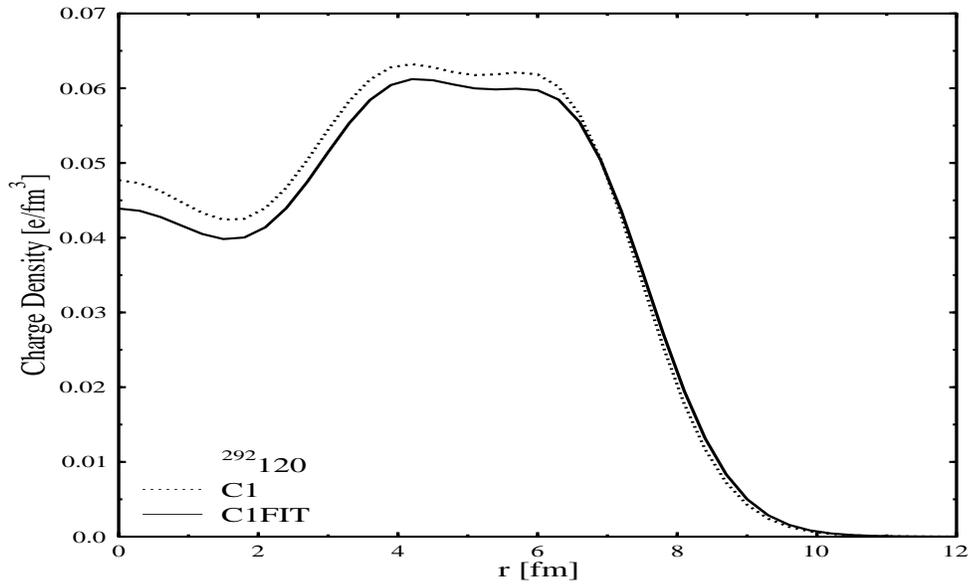, height=10cm ,width=16cm}
\caption{\label{chrgd}Charge Density Distribution of
$Z=120$, $N=172$}
\end{figure}
\begin{figure}[h]
\vspace*{-0.3cm}
\epsfig{file=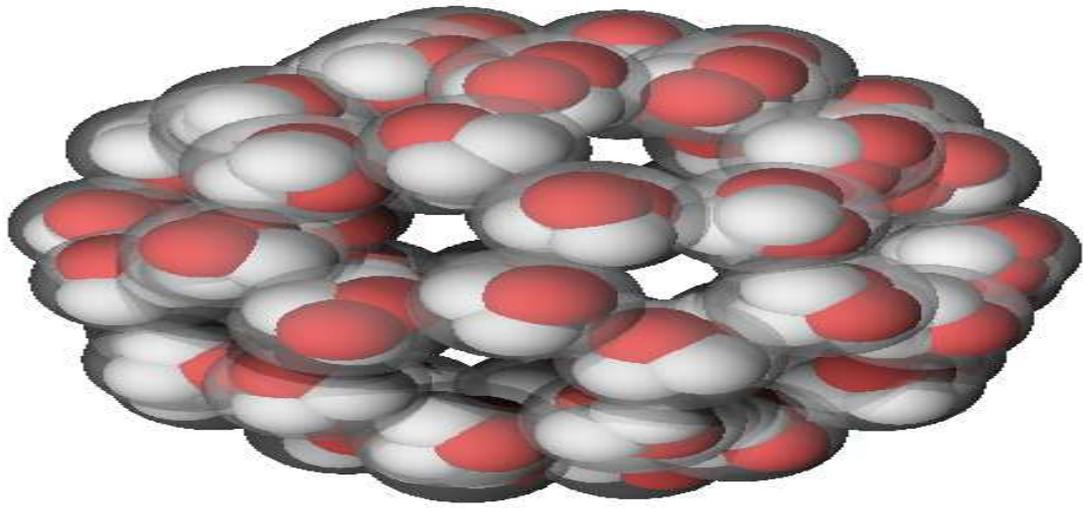, height=10cm ,width=16cm}
\caption{\label{bucky}Alpha-Cluster with 60 $\alpha$-Particles}
\end{figure}

\vspace{0.5cm}
Reasonable descriptions of nuclei directly derived from QCD 
models are still not in sight. 
However, one can investigate effective models that incorporate
basic symmetries of QCD, describing hadronic 
properties as well as nuclear matter and finite nuclei
\cite{hei94,paper3}.
The approach discussed here \cite{paper3} builds on chiral 
$SU(3)_{\mathrm L}\times SU(3)_{\mathrm R}$
symmetry and broken scale-invariance.
The model incorporates relativistic baryonic and mesonic
degrees of freedom
(nucleons, hyperons, spin $\frac{3}{2}$ baryons, 
the spin-0 and spin-1 $SU(3)$-multiplets \cite{paper3}).
The hadron masses are generated dynamically via
spontaneous symmetry breaking.
The masses of the nucleons are generated 
by the interactions with a non-strange scalar field
$\sigma\sim\langle\overline{u}u+\overline{d}d\rangle$ and
a strange condensate
$\zeta\sim \langle\overline{s}s\rangle$ as 
\begin{equation}
m_{\mathrm N}=g_{{\mathrm N}\sigma}\sigma+g_{{\mathrm N}\zeta}\zeta.
\end{equation}
Other baryonic masses are generated in the same manner.
The pseudoscalar mesons obtain their masses by explicit symmetry 
breaking.\\
Our present investigation of superheavy nuclei uses three 
different sets of parameters. The calculations are performed in
spherical symmetry adopting the mean-field approximation.
Free parameters are fixed to vacuum properties of hadrons and
ground state properties of infinite nuclear matter (Set C1) 
and via 
a $\chi^2$-fit to properties of the spherical nuclei 
$^{16}$O, $^{40}$Ca, $^{48}$Ca, $^{58}$Ni, $^{90}$Zr, $^{112}$Sn,
$^{124}$Sn and $^{208}$Pb (Set C1fit).\\ 
The $\chi^2$-function is given by
\begin{equation}
\chi^2=\sum\limits_{n}\left(\frac{{\cal O}^{\mathrm exp}_n
- {\cal O}^{\mathrm theo}_n}{\Delta{\cal O}_n}\right)^2,
\end{equation}
where ${\cal O}^{\mathrm{exp}}_n$ is the experimental value of 
the observable ${\cal O}_n$, while ${\cal O}^{\mathrm{theo}}_n$ is 
its calculated value. $\Delta{\cal O}_n$ is a weight
given by the experimental error. However, because all observables 
are known experimentally to much better accuracy than provided by 
the model, we have decided to use the 
weights to make the fits comparable 
to other calculations. The observables used
for the fits to the nuclei are the binding 
energy $E_{\mathrm B}$, the 
surface thickness $\sigma$ and the charge diffraction radius 
$R_{\mathrm{diff}}$. For extrapolation to heavy 
nuclei one should replace $^{16}$O by 
$^{264}$Hs as done in the parameter set C1hs.
All parameter sets give reasonable results in the region 
of known nuclei. This is a remarkable result particularly for 
the fit to hadron properties and nuclear matter (C1), which does not 
contain any information about finite nuclei.
The fit to finite nuclei can be used to improve the $\chi^2$.
All parameter sets give the correct shell closures
for magic nuclei up to 82 protons and 126 neutrons.\\
Several experiments have recently attempted to produce 
nuclei with proton numbers beyond the element $Z=112$ that was 
found at GSI \cite{hofmann}. 
Experiments report observation of the nucleus 
$^{293}118_{175}$ \cite{ninov}. 
Also $Z=114$ has been reported \cite{ogan}.\\
The most important theoretical question is 
where new shell closures are to be expected.
These nuclei could have very long lifetimes (minutes to years).
This topic has been extensively investigated in Walecka type models
\cite{rutz} predicting $Z=114$ and $Z$=120 as next shell closure.
Let us now turn to the predictions of the chiral model.
Figure \ref{gap} shows the finite-nucleus calculation 
(in the chiral model) of the two-proton gap 
\begin{equation}
\delta_{\mathrm 2p}(Z,N)=S_{\mathrm 2p}(Z,N)-S_{\mathrm 2p}(Z+2,N)
\end{equation}
with the 2p-separation energy $S_{\mathrm 2p}$ defined as
\begin{equation}
\label{sepen}
S_{\mathrm 2p}(Z,N)=E_{\mathrm B}(Z-2,N)-E_{\mathrm B}(Z,N)
\end{equation}
The calculation was done for a nucleus with $N=172$.
The signature for a shell closure 
is a pronounced peak in the two-nucleon gap.
One can see that all three parameter sets show a shell closure 
at $Z=120$. At $Z=114$ a small peak is also visible.
It is interesting to see that the weak peak at $Z=114$ is 
suppressed, if the parameter set includes the binding energy of
$^{265}$Hs (C1hs),
thus taking into account a known heavy nucleus in the 
vicinity of $Z\sim$ 114.
This seems to suggest that $Z=114$ is not a magic number
(C1hs should give the best prediction for superheavy nuclei).
This result has recently also been obtained in Walecka-type models
-- the early famous $Z$=114 predictions are attributed to an 
incorrect spin-orbit force in nonrelativistic models \cite{rutz}.
The two-nucleon separation energy (\ref{sepen}) shows that for a 
neutron number $N=172$ the nucleus with $Z=120$ is beyond the 
dripline ($S_{\mathrm 2p}<0$) in the calculation with 
the "nuclear matter" fit parameters C1 
(Figure \ref{sep}). 
However for the parameter sets
C1fit and C1hs (adjusted to properties of finite nuclei)
$^{292}120$ is a bound doubly magic nucleus,
clearly above the dripline.
And for 184 neutrons, $Z=120$ yields $S_{\mathrm 2p}>0$ for all 
three parameter sets. $N=172$, 184 and 198 are 
closed neutron shells.
Figure \ref{alpha} shows the predicted $\alpha$ 
particle decay chain of the nucleus 
$Z=120$ and $N=172$.
The $\alpha$ energy decreases slightly towards the 
lighter daughter nuclei and drops to about 4 Mev 
for $Z=106$.
This overall trend is in qualitative agreement with the 
experimental finding of nearly constant $\alpha$-energies 
for $Z$=118, $N$=175 \cite{ninov}.
One should keep in mind that this calculations are performed
under the assumption of spherical symmetry which is probably
not fulfilled for this nuclei, except $Z=120$ which shall 
be magic.\\
The predicted charge distribution of $^{292}120$ is shown 
in figure \ref{chrgd}.
Note its strong depletion in the center
of the nucleus. The same effect is seen in 
Walecka model calculations \cite{bender}.
Hence one may speculate that the superheavy nuclei 
around $Z$=120 exhibit a Fullerene, 
bucky-ball structure built of 60 
$\alpha$-particles and $\approx$ 60 neutrons 
(See figure \ref{bucky}).
The length of the links between the 
$\alpha$-particles in a Fulleren-type structure for $Z$=120 
(radius 8.0 fm) is about 3.5 fm. 
This is surprisingly close to twice the radius
of $^4$He ($R_{\mathrm{diff}} = 1.68$ fm).
Such a structure naturally exhibits a depleted charge 
density in the 
center of the nucleus.
\section*{Acknowledgments}
This work was supported by GSI, BMBF, DFG and the Josef Buchmann 
Stiftung.
We thank Henning Weber for his help with figure 5.


\end{document}